\documentstyle[twoside,epsf]{article}
\def\ben{\begin{eqnarray}}
\def\enn{\end{eqnarray}}
\def\bea{\begin{eqnarray}}
\def\eea{\end{eqnarray}}
\def\be{\begin{equation}}
\def\ee{\end{equation}}
\def\ov{\over\displaystyle\strut}
\def\dst{\displaystyle\phantom{|}}
\def\l({\left(}
\def\r){\right)}

\def\ove#1{\overline{#1}}
\def\bk{{\bf K}}
\def\bdk{{\bf \Delta k}}
\def\rl{R_L^2}
\def\ro{R_{out}^2}
\def\rs{R_{side}^2}
\def\rol{R_{out,L}^2}
\def\rpa{R_{\parallel}^2 }
\def\rpe{R_{\perp}^2 }
\def\rta{R_{\tau}^2 }

\begin{document}
%\rightline{DRAFT 2.0}
\vspace*{10mm}
\noindent
{\Large\bf
	On source parameters from particle correlations \\ and spectra
}
\\[3mm]
\def\rightmark{ On source parameters}
\def\leftmark{A. Ster, T. Cs{\"o}rg\H o and J. Beier}
\hspace*{6.327mm}
{\large\lineskip .75em
A. Ster\footnote{E-mails: \it ster@rmki.kfki.hu, csorgo@sunserv.kfki.hu, 
jochen.beier@ep4.phy.uni-bayreuth.de}, 
T. Cs{\"o}rg\H o$^{2}$ and J. Beier$^{3}$ 
}\\[2.812mm]
\hspace*{6.327mm}
\begin{minipage}[t]{11.2cm}
\hspace*{-8pt}$^1$ MTA KFKI Research Institute for Materials Science \\
H-1525 Budapest 114, POB 49, Hungary  \\
\hspace*{-8pt}$^2$ MTA KFKI RMKI\\
H-1525 Budapest 114, POB 49, Hungary  \\
\hspace*{-8pt}$^3$ Department for Experimental Physics IV, \\
	University of Bayreuth, Universit\"atstrasse 30, \\
	D - 95440 Bayreuth,  Germany\\
\\[5.624mm]\noindent
{\bf Abstract}
	Analytic and numeric approximations are studied in detail for
	a hydrodynamic parameterization of single-particle spectra and 
	two-particle correlation functions in high energy 
	hadron-proton and heavy ion reactions.
	Two very different sets of model parameters
	are shown to result in similarly shaped 
	correlation functions and single particle spectra
	in a  rather large region of the momentum space.
	However, the {\it absolute normalization} of the single-particle 
	spectra is found to be highly sensitive to the choice of 
	the model parameters. For data fitting the analytic formulas
	are re-phrased in terms of parameters of direct physical meaning,
	like mean transverse flow. The difference between the analytic and
	numeric approximations are determined as an analytic function of
	source parameters. 
\end{minipage}

\section{Introduction}
	In 1994-95, a series of papers were written by the Buda-Lund
	collaboration on the study of particle correlations and single-particle
	spectra for non-relativistic, three-dimensionally expanding
	as well as for relativistic, one-dimensionally expanding 	
	or three-dimensionally expanding finite systems \cite{plb94,plb95,3d}.
	In these papers, it has been emphasized for the first time,
	that observation of the ``true" sizes of particle sources
	is possible only if the single-particle spectra and
	the two-particle correlation functions are simultaneously
	analyzed. The reason was also given: the HBT radii 
	(effective sizes measured by correlation techniques)
	were found to be dominated by the shorter of the
	geometrical and the thermally induced lenght-scales,
	while the width of the rapidity distribution or the
	slope of the transverse mass distribution is found to
	be dominated by the longer of the geometrical and the
	thermal scales. The appearance of the thermal lenght-scales
	is related to flow and temperature gradients, i.e. to the
	change of the mean momentum of the emitted particles
	with changing the coordinates of the particle emission.
	Within a thermal radius, these changes are not bigger than
	the width of the local momentum distribution.

	This important effect is presently being re-discovered by
	various other groups, as a consequence of the
	emerging simultaneous analysis of particle correlations
	and spectra, proposed first in 1994 by the Buda-Lund
	collaboration. The question arizes: Is it possible to
	uniquely determine the geometrical source radii
	(``true" source sizes) from a simultaneous analysis
	of particle correlations and spectra? This question is
        basically the same as the analogous question in the momentum
	space: Is it possible to uniquely determine the freeze-out
	temperature and the transverse flow from a simultaneous 
	analysis of particle spectra and correlations?

	We prove by an example that if the absolute normalization
	of particle spectra is {\it not} given, then it may be impossible
	to select from among different minima based only on the 
	shape of the single-particle momentum-distribution and
	on the two-particle correlation functions.	

	We perform the analysis with the help of an analytically
	as well as numerically well studied model, the hydrodynamical
	parameterization of the Buda-Lund collaboration \cite{3d}.	
	The domain of applicability of the analytical approximations 
	is determined numerically in ref.~\cite{3dnum} for this model. 
	In ref.~\cite{na22}, this model is shown to describe the 
	single-particle spectra and the two-particle correlations at 
	$(\pi/K) + p$ reactions at CERN SPS simultaneously.
	
	The  same model was tested in refs.~\cite{3dqm,3ds96} 
	against the {\it preliminary} NA44 data on $S + Pb$ reactions at
	CERN SPS, however, it was shown there that it is 
	difficult to find unique, reliable values of the fit parameters. 
	The sources of the difficulty are the lack of absolute normalization
	of spectra and the experimental difficulty of proper estimate
	of systematic errors. 
	We know from  earlier fitting of this model to 
	spectra without absolute normalization~\cite{3ds96}, that the 
	final results are rather sensitive to errors and normalizations.
	In fact, we  compare here the two physically different minima,
	found by fitting the Buda-Lund hydro model of ref.~\cite{3d}
	to NA44 preliminary data on $S+Pb$ central reactions at 200 AGeV.
	In ref.~\cite{3dcf98} results were reported on 	
	fitting simultaneously the recently obtained
	absolutely normalized but still preliminary
	particle spectra and final correlations data for 	
	the same reaction as analysed in ref.~\cite{3ds96}.

	In the next section the hydrodynamic model is presented, along
	with a new reparameterization of the basic formulas. 
	An approximate  analytic solution to this
	model is formulated in a new manner. 
	For  a comparision, a numerical approximation method is also
	schemed up. In the subsequent section, radius parameters
	and single particle spectra  are calculated using the analytic 
	and numeric methods. The transverse mass and the
	rapidity dependence of the   results are
	shown for a substantial range of momentum space. 
	The results are also used to make estimations to the
	systematic errors introduced by the particular approximations.
	Finally, we summarize and emphasize the importance of 
	the experimental determination of the absolutely normalized
	single particle spectra. 

\section{The model and its re-parameterization}

	The hydrodynamic model of ref.~\cite{3d} is 
	briefly recapitulated below, in a general form.
	The analytic results are then reformulated with new notation.
	A numerical evaluation scheme is also  summarized afterwards.	

\subsection{\it The model}
	
	The Buda-Lund model \cite{3d} model makes 
	a difference between the central (core) and the outskirts (halo)
	regions of high energy reactions.
	The pions that are emitted from the core consist of 
	two types:
	a) They could be emitted directly from the hadronization of wounded,
	string-like nucleons, rescattering with a typical 1 fm/c
	scattering time as they flow outwards.
	b) Alternatively, they could be produced from the decays of
	short-lived resonances such as $\rho$, $N^*$, $\Delta$ or $K^*$,
	whose decay time is also of the order of 1-2 fm/c.
	This core region of the particle source is resolvable by Bose-Einstein
	correlation measurements. 
	In contrast, the halo region consists of decay products of long-lived
	resonances such as the $\omega$, $\eta$, $\eta'$ and $K^0_S$, 
	whose decay time is greater than 20 fm/c. This halo is not resolvable
	by Bose-Einstein measurements with the present techniques,
	however, it is affecting the Bose-Einstein correlation functions
	by suppressing their strength.

	In general, the following emission function 
	$S_c(x,p)$ applies to a hydrodynamically evolving core of 
	particle source:
\ben
        S_{c}(x,p) \, d^4 x & = & {\dst  g \ov (2 \pi)^3} \,
        {\dst d^4 \Sigma^{\mu}(x) p_{\mu} \ov
        \exp\l({\dst  u^{\mu}(x)p_{\mu} \ov  T(x)} -
        {\dst \mu(x) \ov  T(x)}\r) + s }, \label{e:sc}
\enn
	where the subscript $_c$ refers to the core,
	the factor $ d^4 \Sigma^{\mu}(x) p_{\mu} $ 
	describes the flux of particles through a
	finite, narrow layer of freeze-out hypersurfaces.
	The statistics is encoded by $s$, Bose-Einstein statistics
	corresponds to $s = -1$, Boltzmann approximation to 
	$s = 0$ while the Fermi-Dirac statistics corresponds 
	to $s = +1$.
	The four-momentum reads as $p = p^{\mu} = (E_{\bf p}, {\bf p})$. 
	The four-coordinate vector reads as $x = x^{\mu} = (t,r_x,r_y,r_z)$. 
	For cylindrically symmetric, three-dimensionally expanding, 
	finite systems it is assumed that any of these
	layers can be labelled  by a unique value of
	$ \tau = \sqrt{t^2 - r_z^2} $, and the random variable $\tau$
	is characterized by a probability distribution, such that
\ben
         d^4 \Sigma^{\mu}(x) p_{\mu} & = &
      m_t \cosh[\eta - y]  \, H(\tau) d\tau \, \tau_0 d\eta \, dr_x \, dr_y.
\enn
	Here $m_t = \sqrt{m^2 + p_x^2 + p_y^2}$ stands for the transverse
	mass, the rapidity $y$ and the space-time rapidity $\eta$ are 
	defined as $y = 0.5 \log\left[(E+ p_z) / (E - p_z)\right]$ and
	$\eta = 0.5 \log\left[(t + r_z) / (t - r_z)\right]$ and
        the  duration of particle emission is characterized by
        $H(\tau) \propto \exp(-(\tau-\tau_0)^2
        /(2 \Delta\tau^2))$.  Here
        $\tau_0$ is the mean emission time,
        $\Delta \tau$ is the duration of the emission in (proper) time.
        The four-velocity  and the local temperature and density profile
         of the expanding matter is given by
\ben
        u^{\mu}(x) & = & \l( \cosh[\eta] \cosh[\eta_t],
        \, \sinh[\eta_t] {\dst r_x \ov r_t},
        \, \sinh[\eta_t] {\dst r_y \ov r_t},
        \, \sinh[\eta] \cosh[\eta_t] \r), \\
	\sinh[\eta_t] & = & b {r_t \ov \tau_0}, 
	\qquad r_t \,\,  = \,\, \sqrt{r_x^2 + r_y^2},
\enn
	assuming a linear transverse flow profile.
	The inverse temperature profile is characterized by the 
	central value and its variance in transverse and temporal
	direction, and we assume a Gaussian shape of the local density
	distribution:
\ben
{\dst 1 \ov T(x)} & =  &
	{\dst 1 \ov T_0 } \,\,
	\left( 1 + a^2 \, {\dst  r_t^2 \ov 2 \tau_0^2} \right) \,
	\left( 1 + d^2 \, {\dst (\tau - \tau_0)^2 \ov 2 \tau_0^2  } 
	\right),
	\\
	{\dst \mu(x) \ov T(x) }  & = &  {\dst \mu_0 \ov T_0} -
        { \dst r_x^2 + r_y^2 \ov 2 R_G^2}
        -{ \dst (\eta - y_0)^2 \ov 2 \Delta \eta^2 }, \label{e:mu}
\enn
	where $\mu(x)$ is the chemical potential and $T(x)$ is the local
        temperature characterizing the particle emission. 
%	Note that the
% 	strength of the transverse changes of the temperature profile,
%	the gradient of the transverse flow and the 
%	strength of the temporal  changes of the temperature
%	profile are controlled by the dimensionless parameters
%	$a$, $b$ and $d$, respectively.

\subsection{\it Core/halo correction}
	
        The effective intercept parameter 	
	$\lambda_*(y,m_t)$ of the Bose-Einstein correlation function 
	measures the fraction of pions from the core versus the 
	total number of pions at a given value of ${\bf p}$, when 
	interpreted in the core/halo picture ~\cite{chalom,chalo,nhalo}. 
	With this factor the total invariant spectrum
	in $y$ rapidity and transverse mass $m_t$ follows as
\ben
        {\dst d^2 n\ov dy \, dm_t^2 } & = &
                        {\dst 1 \ov \sqrt{\lambda_*} }
                        {\dst d^2 n_{c} \ov dy \, dm_t^2 } \,\, =  \,\,
		{\dst 1 \ov \pi \, \sqrt{\lambda_*} } \,
                {\int S_c(x,p) \, d^4 x}. 
\enn
	The momentum dependence of $\lambda_*$ parameter 
	is to be measured by the experimental collaborations.
	The experimental determination of $\lambda_*(p)$ is
	\underline{very important} not only because it gives a
	measure of the contribution of the core to the total number
	of particles at a given momentum, but also as it provides 
	a measure of the mean transverse flow and a new signal of
	partial $U_A(1)$ symmetry restoration \cite{xx}. 
	Therefore it is strongly recommended that
	experiments report this $\lambda_*({\bf p})$ 
	parameter of the Bose-Einstein correlation function 
	and not just present partial fit results,
	like the momentum dependence of the radius parameters.

\subsection{\it Re-parameterization}

	The original version of the core model  contains 3 dimensionless
	parameters, $a$, $b$ and $d$, that control the
	transverse decrease of the temperature field, the strength of the
	(linear) transverse flow profile and the temporal changes of the 
	temperature field, respectively, keeping only the mean and the 
	variances of the inverse temperature distributions.
	These are very useful in obtaining simple formulas, however,
	they make the interpretation of the fit results less transparent.
	Hence we re-express them with new parameters with more
	direct physical meaning.
	
	The surface temperature is introduced as 
	$T_r = T (r_x = r_y = R_G, \tau = \tau_0)$ 
	and the ``post-freeze-out" temperature denotes
	the local temperature after most of the 
	freeze-out process is over,  
	$T_t = T(r_x = r_y = 0; \tau = \tau_0 + \sqrt{2} \Delta\tau)$. 
	Here $R_G$ stands for the transverse geometrical radius 
	of the source, $\tau_0$ denotes
	the mean freeze-out time, $\Delta \tau$ is the duration of the
	particle emission and we denote the temperature field by $T(x)$.
	The central temperature at mean freeze-out time is denoted by
	$T_0 = T(r_x = r_y = 0; \tau = \tau_0)$.

	Then the relative transverse and temporal temperature decrease
	 can be introduced as
\ben
	\langle {\Delta T / T}\rangle_r & = & {T_0 - T_r \ov T_r}, \label{e:tr} \\
	\langle {\Delta T / T}\rangle_t & = & {T_0 - T_t \ov T_t}, \label{e:tt}
\enn
	and it is worthwhile to introduce the mean transverse flow
	as the transverse flow at the geometrical radius as
\be
	\langle u_t \rangle = b \, {R_G \over \tau_0}. \label{e:ut}
\ee
	The dimensionless model parameters can thus be expressed
	with these new, physically more straightforward
	parameters as
\ben
   {a^2} & = & {\tau_0^2 \over R_G^2} \,\,
			\langle {\Delta T / T}\rangle_r, \\	
   {b } & = & { \tau_0 \over R_G }\,\, \langle u_t \rangle,   \\
   {d^2 } & = & {\tau_0^2 \over \Delta \tau^2} \,\,
			\langle {\Delta T /T}\rangle_t.
\enn

	Note, that eqs.~(\ref{e:tr}) and (\ref{e:ut}) were introduced earlier
	in ref.~\cite{na22} also to simplify the interpretation of data
	fitting. We present the complete re-parameterization herewith,
	including egs.~(\ref{e:tt}) and the re-parameterization of both the
	radius parameters and single-particle spectra.

\subsection{\it Analytic approximations}

	In Ref.~\cite{3d}, the Boltzmann
	approximation to the above emission function 
	was evaluated in an analytical manner, 
	applying approximations around the saddle point of the
        emission function. The resulting formulas express the
        Invariant Momentum Distribution (IMD) and the Bose-Einstein
	correlation function (BECF). Now we re-express the formulas
	given in ref.~\cite{3d} with the help of our new parameters.

	The particle spectra can be expressed in the following simple form: 
\ben
	N({\bf p})	& = &
		{\dst g \ov (2 \pi)^3 } \,
		\overline{E} \,
		\overline{V} \,
		\overline{C} \,
		\exp\left( 
		- { p \cdot u(\overline{x}) - \mu(\overline{x})
		\ov T(\overline{x}) }
		\right),
		\label{e:invn} \\
	\overline{E} & =& 
		m_t \cosh(\overline{\eta}), 
		\label{e:inve}  \\
	{\overline{V}} & = &
		( 2 \pi )^{(3/2)}\, \overline{R_{\parallel}} \,
				    \overline{R_{\perp}}^2 
		\, {\dst \overline{\Delta\tau} \,
		\ov {\Delta\tau} }, 
			\label{e:invv}
		\\ 
	\overline{C} \, & = & 
		{\dst 1 \ov \sqrt{\lambda_*} } \,
		\exp\left( {\dst \ove{\Delta \eta}^2 \ov 2} \right).
\enn 
	Here the quantity $\ove{x}$ stands for the average
	value of the space-time four-vector parameterized by
	$(\ove{\tau},\ove{\eta},\ove{r_{x}},\ove{r_{y}})$,
	denoting longitudinal proper-time, space-time rapidity and
	transverse directions. These values are given as
\ben
        \overline{\tau} & = & \tau_0, \\
        \overline{\eta} & = &
                {\dst y_0 - y \ov 1 + \Delta\eta^2 {\dst m_t \ov T_0 }  },\\
        \overline{r_{x}} & = & \langle u_t \rangle  R_G
                {\dst p_t \ov T_0 +
                \overline{E} \left( \langle u_t\rangle +
                        \langle {\dst \Delta T / T}\rangle_r
                        \right) },       \\
        \overline{r_{y}} & =  0.
\enn
	$\overline{E}$ denotes the
	energy of a particle from the center of particle emission,
	measured in the Longitudinal Center of Mass System
	(LCMS) frame.
	The effective volume is denoted by $\ove{V}$, see below for 
	details, and the correction factor $\ove{C}$ takes into
	account the effects of long-lived resonances and the 
	deviation of the saddle-point result from the more 
	possible naive expectation, which would be the same expression
	with $\ove{C} = 1$.
	The notation $\overline{a}$ denotes
	an invariant quantity $a$, that depends on $y - y_0$, $m_t$,
	$T$  and the other parameters of the model
	in a boost-invariant manner. In ref.~\cite{3d} this 
	was denoted by $\ove{a_*}$, for example, $\ove{\Delta\eta}$ in the
	present paper was denoted by $\ove{\Delta\eta_*}$. 
	The average invariant volume of particle emission 
	in eqs.~(\ref{e:invn},\ref{e:invv}) is
	given by ${\overline{V}}$ that is in our case the 
	product of the average (momentum-dependent) transverse
	area $(2 \pi \overline{R_{\perp}}^2 )$, 
	the average (momentum-dependent) longitudinal source size
	$  ( 2 \pi)^{1/2} \overline{R_{\parallel}}$.
	The averaging of the size of the effective volume 
	over the duration of particle emission is
	expressed by the factor 
	${\dst \overline{\Delta\tau} \ov {\Delta\tau} }$.
	These quantities read as
\ben
	\ove{\Delta\tau}^2 & = & 
		{\dst \Delta \tau^2 \ov
		1 + \langle { {\small \Delta T / T} }\rangle_r 
		{\dst \ove{E} \ov T_0}  }, \\
	\ove{\Delta\eta}^2 & = & 
		{\dst \Delta\eta^2 \ov
		1 +  \Delta\eta^2 {\dst \ove{E} \ov T_0} },\\
	\overline{R_{\parallel}}^2 & = &
		\overline{\tau}^2 \, \overline{\Delta\eta}^2, \\
	\overline{R_{\perp}}^2 & = & 
		{\dst R_G^2 \ov 
		1 + \left( \langle u_t \rangle^2	
		+ \langle { {\small \Delta T / T } }\rangle_r \right)
		{\dst \ove{E} \ov T_0}  }. \\
\enn
	This completes the specification of the shape of particle spectrum.
	These results correspond to the equations given in ref.~\cite{3d}
	although they are re-expressed with new combination of the
	variables.
	Please note that the Boltzmann-factor can be expressed
	approximately as
\ben
		\exp\left( - { p \cdot u(\overline{x}) - \mu(\overline{x})
		\ov T(\overline{x}) } \right) 
			& \simeq &
		\exp\left[ 
		{\dst \mu_0 \ov T_0} 
		- { \dst (y - y_0)^2  \ov 2(\Delta\eta^2 + T_0 / m_t)} \right]
		\, \exp\left[ -{\dst m_t \ov T_0} \right] \times \nonumber \\
		\null & \null &
		\, \times 
		\exp\left[ {\dst \langle u_t \rangle^2 (m_t^2 - m^2) \ov
		2 T_0 \left[ T_0 +
		 m_t ( \langle u_t \rangle^2 + \langle {
		{\small \Delta T / T}} \rangle_r ) \right]
		} \right]. 
\enn 
	The HBT radius parameters were evaluated in ref.~\cite{3d}
	in the following way:
	the space-time rapidity $\eta_s$ was defined as the solution
	of the equation
\be
	{\partial S(\eta_s)\over \partial \eta} = 0.
\ee
	The Longitudinal Saddle Point System (LSPS) was introduced
	as the frame where $\eta_s = 0$. At the so-called saddle-point
	one has
\be
	{\partial S \over \partial \tau} = 
	{\partial S \over \partial \eta} =
	{\partial S \over \partial r_x} =
	{\partial S \over \partial r_y} = 0.
\ee
	This resulted in a set of transcendental equations for the position
	of the saddle-point. These equations for the positions were solved
	approximatelly with the help of an expansion of the transcendental
	equation in terms of small parameters, like the deviation of $\eta_s$
	from the mean rapidity of the pair.
	The calculation resulted in the following value for
	$\eta_s$:
\be
	\eta_s = y + {\dst y_0 - y \ov 1 + \Delta \eta^2 ({m_t \over T_0} -1) }. 
\ee
	Hence the relative rapidity of the LSPS frame as compared to
	the LCMS frame 
	is
\be
	\eta_s^L = {\dst y_0 - y \ov 1 + \Delta \eta^2 ({m_t \over T_0} -1) }.
\ee
	Note that LCMS is the 
	frame where the rapidity
	belonging to the mean momentum of the pair vanishes, 
	$ y = 0$~\cite{lcms}.
	In ref.~\cite{3d} this quantity was denoted by the slightly more
	complicated notation $\eta^{LCMS}_s = \eta_s^{L} 	$,
	and the maximum of the Boltzmann factor,
	which in our present notation 
	reads as $\overline{\eta} =  \overline{\eta_s}$ 
	and stands for a modified saddle-point, used for the 
	calculation of the particle spectra in ref.~\cite{3d}.
	Note that $\eta_s^L$ may deviate from $\overline{\eta}$
	substantially at low values of $m_t$, especially if
	$T > m$.

	This frame, defined by the maximum of the Boltzmann factor, 
	given by $\overline{\eta}$ in the LCMS,
	 plays a key role in the calculations.
	Hence this frame, introduced in ref.~\cite{3d}
	without a name, deserves a name. We suggest Longitudinal
	Boltzmann Center System (LBCS). In general, this LBCS frame
	is defined by the solution of the 
\be
	{\partial f_B(\overline{\eta})\over \partial \eta} = 0 
\ee
	equation, where 
	$f_B = \exp\left(- [p u(x) - \mu(x)]/T(x) \right) $	
	stands for the Boltzmann factor only, but does not include
	the Cooper-Frye flux term.
	A more detailed comparision between 
	the LBCS and LSPS frames is in preparation for a  
	separate publication~\cite{csl}.
	
	In ref.~\cite{3d} the spectrum was evaluated in the LBCS frame,
	while the correlation functions in the LSPS frame.
	Let us recapitulate the results for the two-particle
	correlation functions:
	The mean momentum is denoted by $K = (p_1 + p_2)/2$,
	its components are $(K_0, K_L, K_t,0)$ in the 
	$(t,r_z,r_x,r_y)$ reference frame.
	The transverse velocity $\beta_t$ reads as
\be
	\beta_t  = 
		{K_t\over K_0} \, = \, {K_t \over  M_t }
		{ 1 \over \cosh(y)} = {V_t \over \cosh(y) }.
\ee
	The effective source parameters are obtained as
\bea
	{1 \over {\Delta\eta_*^2}} & = & {1 \over 
		{\overline{\Delta\eta}^2( \overline{\eta}
		 \rightarrow \eta_s^L )}}
		- {1 \over \cosh^2(\eta_s^L)}, \\
		\rpa
		& =  & \tau_s^2 \Delta\eta_*^2
		\, = \, 
		\tau_s^2 \overline{\Delta\eta}^2
		( \overline{\eta} \rightarrow \eta_s^L ), 
		\label{e:rpa}\\
	R_{\perp}^2 & = & R_*^2  
		\, = \, 
		\overline{R_{\perp}}^2
		( \overline{\eta} \rightarrow \eta_s^L ),
		\label{e:rperp} \\
	R_{\tau}^2 & = & \Delta\tau_*^2 
		\, = \, 
		\overline{\Delta\tau}^2
		( \overline{\eta} \rightarrow \eta_s^L ).
		\label{e:rtau}
\eea 
	The modified position of the maximal emissivity is given in the
	LCMS frame by $(\tau_s,\eta_s,r_{x,s},0)$, where
\bea
	        r_{x,s} & = & \overline{r_x}
		              (\overline{\eta} \rightarrow \eta_s^L ), \\
		\tau_s  & = & \overline{\tau} = \tau_0.
\eea
	This notation means that, for example, $\Delta\eta_*$ corresponds to
	the function $\overline{\Delta\eta}$ when the variable
	$\overline{\eta}$ is replaced by the quantity
	$\eta_s^L$. Note, that $\Delta\eta_*$ corresponds to the
	effective space-time rapidity width of the particle emission
	function in an LSPS calculation, while $\overline{\Delta\eta}$
	corresponds to the width of the Boltzmann-factor only in the
	LBCS frame. Similar relations hold for $\ove{R}$ and $R_*$,
	$\ove{\Delta\tau}$ and $\Delta\tau_*$.
	The Bose-Einstein correlation functions can be written 
	in the so-called Bertsch-Pratt side-out-long ref. frame
	as
\bea
	C(\Delta k,K) & = & 1 + \lambda_* 
		\exp\left( - \rs Q_{side}^2 
	- \ro Q_{out}^2 - \rl Q_L^2 - \rol Q_L Q_{out} 
			\right), \nonumber \\
	\null & \null & \null \\
	\rs & = & \rpe = R_*^2, \\
	\ro & = & \rs +\delta\ro, \\
	\delta\ro & = & {\dst V_t^2 \ov \cosh^2(y) } 
		\left[ \cosh^2(\eta_s) \rta + \sinh^2(\eta_s) \rpa \right],\\
	\rl & = & {\dst 1 \ov \cosh^2(y)}
	\left[ \cosh^2(\eta_s^L) \rpa + \sinh^2(\eta_s^L) \rta \right],\\
	\rol & = & - {\dst V_t \ov \cosh^2(y) }
		\left[ \cosh(\eta_s) \sinh(\eta_s^L) \rta
		+ \sinh(\eta_s) \cosh(\eta_s^L) \rpa \right]. \label{e:anl}
				\label{e:inl}
\eea
	Here we have utilized the form of equations given in ref.~\cite{3d}
	and the transformation of $K_L$ to $M_t \cosh(y)$ as introduced in
	ref.~\cite{volo}.

	Although the two-particle Bose-Einstein correlation function
	is manifestly covariant~\cite{volo}, its Bertsch-Pratt
	parameterization is frame-dependent. A possible covariant
	parameterization~\cite{ykp1,ykp2}, applied recently to the 
	particle interferometry by the Regensburg group~\cite{ykp3,ykp4}.
	We find that the simplest possible covariant generalization
	is not exactly the  YKP parameterization, but the formulation
	given by the Buda-Lund collaboration in ref.~\cite{3d},
	see especially eqs.~(44) and (21-26) of ref.~\cite{3d}.
	Here we repeat only the results after Gaussian approximation
	to the source function:
\begin{eqnarray}
	C(\Delta k, K)\! & = & 1 + \lambda_* \exp( - Q_{\tau}^2 R_{\tau}^2 
				- Q_{\eta}^2 R_{\parallel}^2 - Q_t^2 R_*^2), \\
	Q_t^2\! & = & Q_{side}^2 + Q_{out}^2, \\
	Q_{\tau}\! & = & Q \cdot n(x_s^L) = 
		Q_0 \cosh(\eta_s^L) - Q_z \sinh(\eta_s^L), \\
	Q_{\eta}\! & = &\! \sqrt{ Q \cdot Q - (Q \cdot n(x_s^L) )^2 
			- Q_t^2 } = 
			Q_0 \sinh(\eta_s^L) - Q_z \cosh(\eta_s^L). 
\end{eqnarray}
	where $n(x_s^L) = (\cosh(\eta_s^L),0,0,\sinh(\eta_s^L) ) $ is
	a normal-vector at $x_s$ in LCMS.
	Note, that this formulation is equivalent with the YKP formulation,
	however the correlation function is given by a simple
	purely quadratic form in the present formulation, in
	contrast to the YKP expression, where the invariant relative
	momentum combination $Q_{\eta}$ is written out explicitely in terms
	of its non-invariant components.
	A more detailed comparision between the YKP and the Buda-Lund
	parameterization is being discussed in ref.~\cite{csl}.

\subsection{\it Numeric approximations}

	Approximate single particle spectra and Bose-Einstein correlation
	functions
        can be calculated by numerical integration of equation (1), as well,
        expressing the means and the variances of the 
	hydrodynamically evolving core of particle emission.
	The original method of 'means and variances' was proposed first
	by the Regensburg group in ref. ~\cite{uli}.
	We shall utilize herewith the core/halo corrected version
	of these relations as given recently in ref.~\cite{nicker}.
	The limitation of these approximations is discussed in
	refs.~\cite{3d,nicker}. For example,
	possible double-Gaussian structures or non-Gaussian features 
	are neglected in this approximation.
        The analytic approximation yields Gaussian functions in
        proper time $\tau$ and space-time rapidity $\eta$,
        hence includes  a deviation
        from Gaussian shape in $t$ and $z$. 
	The numerical approximation
        assumes Gaussian forms in $t$ and $z$ that is not so well suited
        to the kinematics of ultra-relativistic reactions as Gaussians
        in $\tau$ and $\eta$. The spectra and the HBT radius parameters
	are defined in the Gaussian 
	core/halo  model approximation as follows:
\bea
	C(\bk,\bdk ) & = & 1 + \lambda_*({\bf K}) \,
	\exp\left(- R^2_{i,j}({\bf K}) 
		{\bf \Delta k}_i {\bf \Delta k}_j \right), 
		\label{e:cch}\\
	\lambda_*({\bf K}) & = & [ N_{\bf c}({\bf K}) / N({\bf K}) ]^2, \\
	R_{i,j}^2({\bf K}) & = &
		 \langle (x_i - \beta_i t) (x_j - \beta_j t) \rangle_{\bf c}
		 - \langle (x_i - \beta_i t)\rangle_{\bf c}
		   \langle (x_j - \beta_j t) \rangle_{\bf c} \ ,
		\label{e:rch} \\
	\langle f(x,{\bf K}) \rangle_{\bf c} & = & 
		\int d^4 x f(x,{\bf K})  S_{\bf c}(x,{\bf K}),
\eea
	where $i,j = side, out$ or $long$ as before, 
	and $S_{\bf c}(x,{\bf K})$ is the emission function 
	that characterizes the central core, as given by 
	eq.~\ref{e:sc}. The mean momentum is defined as
	${\bf K} = 0.5 ({\bf p}_1 + {\bf p}_2)$,
	the relative momentum is given by
	${\bf \Delta k} = {\bf p}_1 - {\bf p}_2$.
	The spectra  of all the bosons  and the spectra of the
	bosons from the core described as 
\bea
	N({\bf p})  & = & \langle 1 \rangle
			 = \int d^4x \left[ S_{\bf c}(x,{\bf p}) 
			+ S_{halo}(x,{\bf p}) \right], \label{e:ntot} \\
	N({\bf p})_{\bf c}  & = & \langle 1 \rangle_{\bf c}
			 = \int d^4x  S_{\bf c}(x,{\bf p} )
			\label{e:nc}. 
\eea
	In this picture, the reduction of the 
	intercept parameter is the only effect  on the correlation
	function that stems from
	the halo, the variances of the core
	correspond to the Gaussian core/halo model radii of the 
	measured correlation function. 
	Although the above expressions
	are formally similar to the original version of
	Gaussian model-independent radii of ref.~\cite{uli},
	they cannot be obtained with an expansion around $Q = 0$, 
	as  they correspond to a large $Q$ expansion
	of the Bose-Einstein correlation function~\cite{darius,nicker}.

\section{Calculating observables}

	Both analytic and numeric approximations were used to calculate
	and show the momentum dependence of the observables
	from the hydrodynamical model of ref.~\cite{3d}. 
	These observables are the effective radius parameters
	(HBT radii), the shape and the slope of the single-particle 
	spectra. Figures are drawn for two very 
	different sets of source parameters (or model parameters).
	To recall, the particle source is characterized by
	the means and the variances of the density distribution,
	the inverse temperature distribution and a linear flow.
	 This yields 9 free parameters, 
	$\mu_0, T_{0}$, $\tau_{0}$,  $R_{G}$, $\Delta\tau$, $\Delta\eta$, 
	$\langle \Delta T / T\rangle_r$, $\langle u_t \rangle$, 
	$\langle \Delta T / T\rangle_t $, respectively. 

	The examined momentum space is divided into 40x40 sub-intervals
	in $m_t$ and $y$ dimensions, respectively, that allowed for
	fine resolution of the distributions. 
	As a drawback, with such a resolution
	the numeric integration version takes much longer time than the
	analytic one. In the presented case generation of data took ~10 
	hours for one run. Scales on the pictures are kept the same
	for the same kind of distributions
	except for the average emisson rate that differs remarkably
	for the two sets of model parameters.
	The actual parameter set values (with $\mu_0 = 0$) 
	used in the particular 
	calculations are indicated below each drawing
	and they are denoted by names Source Parameters Set 1 
	and Source Parameter Set 2. Parameter Set 1 was obtained
	from a ref.~\cite{3dcf98} , fitting absolutely
	normalized spectra in the NA44 acceptance, while
	Parameter Set 2 was obtanined in ref.~\cite{3ds96},
	fitting unnormalized preliminary particle spectra
	together with correlation data.
	 Note that one can distinguish
	between different model parameters only if the value of 
	$\mu_0$ is known from other observations than the ones
	already exploited in the present analysis.

\subsection{\it Analytic results}

	The two model parameter sets mentioned above were applied
	to the analytic expressions 
	formulated in the previous section by eqs. (\ref{e:invn}) to
	(\ref{e:inl}).
	See Figures 1 and 2 for details of the momentum space distributions
	of the observables.
	Notice the substantial difference of the particle spectra
	for Parameter Set 1 and Parameter Set 2. Also notice the
	deviations of the radius parameters at small
  	$m_t$ and at large relative rapidities to midrapidity for the 
	two different source parameter sets.
	Along with the comparision to the numeric results later this
	reflects the limitations of this kind of approximation.
	See ref.~\cite{csl} for an improved treatement. 

\subsection{\it Numeric results}

	The numerically evaluated HBT radius parameters and single particle
	spectra  are obtained from the 
	eqs. (\ref{e:cch},\ref{e:rch},\ref{e:ntot},\ref{e:nc}),
	utilizing the Boltzmann approximation to the source function 
	of eq.~(\ref{e:sc}). 
        Note that this scheme is not an exact calculation,
        but an approximation in a different way than the analytic approach,
        therefore it is suitable to estimate the systematic errors of the model
        parameters and to cross-check the uniqueness of the minimum in
	fitting the model to experimental data.
	See Figures 3 and 4 for details of HBT radius parameter distributions
	as well as single particle spectra in this approximation scheme.

\subsection{\it Differencies between the analytic and the numeric
	 results}

	The differencies between the observables as calculated from 
	the two sorts of model approximations are presented on Figures 5 and 6.
	From these drawings one can learn the critical ranges
	where the two approximation schemes differ from each other 
	beyond a given tolerance level.

	Figures 5 and 6 show the rapidity and transverse mass dependence
	of the relative deviations between the analytic and the numeric
	approximations. For example, let us consider the top left panel
	of Figure 5. On this panel, the relative deviation of the 
	numerical and analytic approximations for the side radius component is 
	evaluated in the following manner:
	The analytical result for a $(y, m_t)$ bin $i$ is denoted by $a_i$,
	the numerical result is denoted by $n_i$. Then the relative
	deviation between the two approximation schemes is defined as 
\be
	{\delta^2 a_i \ov a^2_i}  = {\dst ( a_i - n_i)^2 \ov a_i^2 },
\ee
	which is plotted on the top left panel for the side radius parameter
	and in subsequent panels 
	for the out, long and cross term, the spectra and the
	slope of the spectra on the subsequent panels.
	This quantity will be understood and estimated in the next
	subsection as a function of some small expansion parameters,
	that are analytically obtainable for any set of source parameters.
	In turn, this result can be utilized to improve the fits of the 
	analytic expressions to measured data.

\subsection{\it Estimating systematic errors of approximations}

	The aim of the present subsection is to analytically understand
	the systematic errors on the analytic approximations that we 
	utilize to evaluate the spectra and the HBT radius parameters.
	
	In a fit with the analytic approximations, the $\chi^2$ of the
	fit is given as
\be 
	\chi^2_a  =  \sum_i {\dst (d_i - a_i)^2 \ov e_i^2},
\ee
	where $d_i$ denotes the measured data point at a given 
	bin, for example, $R_{side} (y_i, m_{t,i}) $, and the experimental
	error on this quantity is given by $e_i$.
	A numeric fit to the same data minimizes the following
\be 
	\chi^2_n =  \sum_i {\dst (d_i - n_i)^2 \ov e_i^2}
\ee
	numeric $\chi^2_n$ distribution. This may have different
	minima for the model parameters from the minima of $\chi^2_a$.
	However, the $\chi^2_n$ distribution can be approximately
	reconstructed from an analytic fit, as follows:
\bea
	\chi^2_n & \simeq & \chi^2_a + \delta\chi^2_a, \\
	\delta\chi^2_a & = & \sum_i {\dst \delta^2 a_i \ov e_i^2}
	\,	= \, 
	\sum_i \delta^2 A_i {\dst a_i^2 \ov e_i^2}, 
\eea
	where $\delta^2 a_i = (a_i - n_i)^2$ is the difference
	between the analytic and numeric result, and can
	be regarded as the systematic error of the analytic approximation, 
	while the relative systematic error of the analytic calculation
	is $\delta^2 A_i = {\delta^2 a_i \ov a_i^2} $. (Keep in mind that
	$a_i$ can be any of the analytically evaluated radius parameters
	or analytic result for the particle spectra).  Our purpose is
	to obtain approximate analytical expressions for the relative
	error of the analytical approximations, $\delta^2 A_i$.
	These quantities are shown in Figures 5 and 6.

	Figures 5 and 6 indicate large relative errors in certain
	regions of the $(y,m_t)$ plane. These regions coincide
	with the regions where the so called small  
	expansion parameters~\cite{3d} of the model
	start to reach values close to 1. The analytic expressions
	for the observables  (HBT radius parameters and single-particle
	spectra) were obtained in ref.~\cite{3d}
	under the condition that the 
	parameters $(\eta_s^{L}, \Delta \eta_*, r_{x,s}/\tau_0)$
	are all much less than 1. This was due to the approximate 
	nature of the solution of the saddle-point equations,
	and the expansion of the transcendental equations in terms
	of these small parameters. 
	Figures 7 and 8 show the $(y,m_t)$ dependence of these 
	small parameters and their squares.
	We observe the expected similarities to the distributions
	of the relative errors $\delta^2 A_i$,
	Figure 5 and 6. The different small parameters
	become large in well separated domains of the momentum
	space (typically below 100MeV and above 1 GeV),
	e.g. $\eta_s^{L}$ at small $m_t$ and large $| y - y_0|$,
	$\Delta \eta_*$ at small $m_t$ and small $| y - y_0|$,
	$r_{x,s}/\tau_0 $ at large $m_t$ independently of $y$.
	As a consequence, the relative errors of the analytic
	approximations for any radius parameter or
	momentum distribution can be parameterized as a 
	linear combination of the squared small parameters:
\be
	\delta^2 A = c^A_1 (\eta_s^L)^2  + c^A_2 \Delta\eta_*^2 +
			c^A_3 (r_{x,s}/\tau_0)^2. 
		\label{e:dai}
\ee
	The three constants $(c^A_1,c^A_2,c^A_3)$ are determined for
	the observables $A = R_{side}$, $R_{out}$, $R_{long}$,
	$R_{out-long}^2$, $N({\bf p})$, $T_{eff}(y,m_t)\,  $.
	Since the small parameters are expressed as a function
	of the model parameters or ``true" source parameters, 
	eqs.~(\ref{e:rpa}-\ref{e:rtau}),
	the desired analytical formula for all 
	the systematic errors of the evaluation of 
	all these 6 observables is given in the 
	form of eq.~(\ref{e:dai}).

	Note that the region, where the analytic expressions are most
	precise, corresponds to a curved region in the $(y,m_t)$ plane,
	that at low $m_t$ starts off-mid-rapidity and at high $m_t$
	shifts to mid-rapidity, in case of pions. This region almost 
	exactly coincides with NA44 acceptance for pions. For heavier
	particles all the 3 small parameters decrease substancially.
	The analytic calculation is thus more precise for heavier 
	particles than for pions.

	For determining the numerical coefficients $c^A_i$,
	the distributions of the differencies on Figures 5 and 6
	the CERN optimizing package MINUIT 
	was used ~\cite{minuit}.  MINUIT finds the
	minimum value of a multi-parameter function and it analyzes
	the shape of the function around the minimum. The fitted
	error distributions in terms of the small parameters are shown on 
	Figures 9 to 12 together with the best estimates of the
	systematic errors with the help of eq.~\ref{e:dai}.
	The coefficients $c^A_i$ of the parameterized systematic error
	distributions along with their errors are shown in Table 1 and Table 2.

	Note that the coefficients $c^A_i$ are found to be only
	weakly dependent on the source parameter sets for most of
	the cases, they are all smaller then the coefficents for 
	Parameter Set 1 multiplied by 2.
	On Figure 11, the fit of $(\delta N(p) / N(p))^2$ reflects the effect
	that above 0.5 MeV the calculation was forced to
	take the amplitudes with low weight due to the fact that
	the absolute values of $N(p)$ are very small in this range.

\section{Conclusion}

	The first combined HBT and spectrum analysis,
	reported by the Buda-Lund collaboration in 1994-95,
	is re-visited here for the purpose of a systematic numerical 
	and analytical evaluation of the  model. 
	We identified the regions in the rapidity - transverse
	mass plane where the analytic approximations and the
	numerical ones deviate from each other. These regions were
	found to coincide with the regions where the small expansion
	parameters of the analytic approximation start to grow
	significantly. The deviation between the analytical and the
	numerical results is characterized by  positive definite
	quadratic polinomials built up from the small parameters.
	We find that the NA44 acceptance is ideally suited for 
	the precise evaluation of the  Buda-Lund model. 

	As a by-product we find that the parameters of the 
	Bose-Einstein correlation functions as well as the shape
	and the slope parameters of the double differential
	invariant momentum distribution are similar within 10\%
	for two physically very different model parameter value sets
	in the momentum space domain where the approximations are valid.
	However, these sets result in a factor of 7 - 10 difference
	in the absolute normalization of the single-particle spectra.
	Hence, it is strongly recommended to publish the experimentally measured
	single-particle spectra with their absolute
	normalization for future CERN and RHIC heavy
	ion experiments,  for as many type of particles as possible,
	as a two-dimentional function of $y, m_t$. Figures 1 to 4
	illustrate that the $y$ and the $m_t$ dependence of
	$N(y,m_t)$ can not be factorized.

	Of course, our {\it example} of two physically different 
	parameter sets resulting in similar correlations and unnormalized
	spectra does not imply that such similarity is achieved for
	{\it any} two different parameter sets. The role of
	each parameter can be investigated, for instance, analytically
	like in ref.~\cite{3d,3dnum}. The typical behaviour that
	one expects is that for different values of the model
	parameters the particle correlations and spectra 
	are different. We would like to remind the
	readers that the existence of certain scaling limiting cases
	was pointed out also in ref.~\cite{3d}, 
	where the dependence of the HBT radii on some of the model
	parameters was analytically shown to vanish in certain
	domains of the parameter space.

	Note that the intercept parameter of the correlation function
	is included in the core/halo correction factor, hence it is
	also strongly recommended to publish the experimental HBT
	results including a momentum-dependent determination
	of  $\lambda_*$, too.

\vskip 10pt
\subsection*{Acknowledgement}
        This work was partially supported by the Hungarian NSF 
        grants OTKA - T016206, T024094 and T026435, by the
	IAESTE exchange programme and by the OTKA-NWO grant
	N25487 as well as by the US- Hungarian Joint Fund 
	MAKA 652/1998, and by a travel grant from the SOROS fundation.
	The authors would like to thank Prof. Bengt L{\"o}rstad for
	a special hospitality at the University of Lund.

% \vfill\eject

\vfill\eject

\begin{center} {\large\bf Table captions} \end{center}

\begin{itemize}
{ \item[\bf Table~1.] 
 Coefficients $c_i^A$ of the parameterized systematic error distributions
 for Source Parameter Set 1.
}

{ \item[\bf Table~2.] 
 Coefficients $c_i^A$ of the parameterized systematic error distributions
 for Source Parameter Set 2.
}
\end{itemize}

\begin{center} {\large\bf Figure captions} \end{center} 

\begin{itemize}
{ \item[\bf Fig.~1.] 
Simultaneous results for particle spectra and HBT radius parameters.
The analytic approximations were utilized to evaluate the model for
Parameter Set 1.
}

{ \item[\bf Fig.~2.]
Simultaneous results for particle spectra and HBT radius parameters.
The analytic approximations were utilized to evaluate the model for
Parameter Set 2.
}

{ \item[\bf Fig.~3.]
Simultaneous results for particle spectra and HBT radius parameters.
The numeric approximations were utilized to evaluate the model for
Parameter Set 1.
}

{ \item[\bf Fig.~4.]
Simultaneous results for particle spectra and HBT radius parameters.
The numeric approximations were utilized to evaluate the model for
Parameter Set 2.
}

{ \item[\bf Fig.~5.]
Relative deviations between the analytic and the numeric approximations
to the HBT parameters and particle spectra
        for Parameter Set 1.
}

{ \item[\bf Fig.~6.]
Relative deviations between the analytic and the numeric approximations 
to the HBT parameters and particle spectra
        for Parameter Set 2.
}

{ \item[\bf Fig.~7.]
Small expansion parameters of the hydrodynamic core model
        for Parameter Set 1.
}

{ \item[\bf Fig.~8.]
Small expansion parameters of the hydrodynamic core model
        for Parameter Set 2.
}

{ \item[\bf Fig.~9.]
Parametrization of the relative deviations between the analytic and the 
numeric approximations to the HBT parameters and particle spectra
        for Parameter Set 1.
}

{ \item[\bf Fig.~10.]
Parametrization of the relative deviations between the analytic and the 
numeric approximations to the HBT parameters and particle spectra
        for Parameter Set 2.
}

{ \item[\bf Fig.~11.]
Parametrization of the relative deviations between the analytic and the 
numeric approximations to the HBT parameters and particle spectra
        for Parameter Set 1.
}

{ \item[\bf Fig.~12.]
Parametrization of the relative deviations between the analytic and the 
numeric approximations to the HBT parameters and particle spectra
        for Parameter Set 2.
}
\end{itemize}

\vfill\eject

\begin{center}
\begin{table}[hc]
\begin{center}
\begin{tabular}{lccc}
\hline
\hline
$~~~~~\delta^2A$   &$c_1^A$ & $c_2^A$ & $c_3^A$ \\
\hline
$(\delta R_{side} / R_{side})^2$
        & 0.015 $\pm$ 0.001 & 0.003 $\pm$ 0.001 & 0.002 $\pm$ 0.001 \\
$(\delta R_{out} / R_{out})^2$
        & 0.013 $\pm$ 0.001 & 0.001 $\pm$ 0.001 & 0.039 $\pm$ 0.001 \\
$(\delta R_{long} / R_{long})^2$
        & 0.027 $\pm$ 0.001 & 0.014 $\pm$ 0.001 & 0.032 $\pm$ 0.001 \\
$(\delta R_{outl}^2 / R_{outl}^2)^2$
        & 0.130 $\pm$ 0.001 & 0.208 $\pm$ 0.001 & 0.001 $\pm$ 0.001 \\
$(\delta N(p) / N(p))^2$
        & 0.014 $\pm$ 0.001 & 0.044 $\pm$ 0.001 & 0.015 $\pm$ 0.001 \\
$(\delta T_{eff} / T_{eff})^2$
        & 0.006 $\pm$ 0.001 & 0.042 $\pm$ 0.001 & 0.001 $\pm$ 0.001 \\
\hline
\hline
\end{tabular}
\end{center}

\hspace{0.5cm}
\begin{minipage}[t]{11.0cm}
\caption{
{\small
% Coefficients $c_i^A$ of the parameterized systematic error distributions
% for Source Parameter Set 1.
}}
\end{minipage}
\end{table}
\end{center}

\begin{center}
\begin{table}[hc]
\begin{center}
\begin{tabular}{lccc}
\hline
\hline
$~~~~~\delta^2A$  &$c_1^A$ & $c_2^A$ & $c_3^A$ \\
\hline
$(\delta R_{side} / R_{side})^2$
        & 0.001 $\pm$ 0.001 & 0.007 $\pm$ 0.001 & 0.003 $\pm$ 0.001 \\
$(\delta R_{out} / R_{out})^2$
        & 0.008 $\pm$ 0.001 & 0.002 $\pm$ 0.001 & 0.001 $\pm$ 0.001 \\
$(\delta R_{long} / R_{long})^2$
        & 0.018 $\pm$ 0.001 & 0.024 $\pm$ 0.001 & 0.031 $\pm$ 0.001 \\
$(\delta R_{outl}^2 / R_{outl}^2)^2$
        & 0.154 $\pm$ 0.003 & 0.398 $\pm$ 0.002 & 0.001 $\pm$ 0.001 \\
$(\delta N(p) / N(p))^2$
        & 0.035 $\pm$ 0.001 & 0.052 $\pm$ 0.001 & 0.001 $\pm$ 0.001 \\
$(\delta T_{eff} / T_{eff})^2$
        & 0.012 $\pm$ 0.001 & 0.042 $\pm$ 0.001 & 0.001 $\pm$ 0.001 \\
\hline
\hline
\end{tabular}
\end{center}

\hspace{0.5cm}
\begin{minipage}[t]{11.0cm}
\caption{
{\small
% Coefficients $c_i^A$ of the parameterized systematic error distributions
% for Source Parameter Set 2.
}}
\end{minipage}

\end{table}
\end{center}

\vfill\eject
\begin{center}
\vspace*{16.5cm}
%\special{psfile=anal1.eps hoffset=-20 voffset=0 hscale=67 vscale=60}
\begin{minipage}[t]{11cm}
\begin{center} {\small {\bf Fig.~1.}
%Simultaneous results for particle spectra and HBT radius parameters.
%The analytic approximations were utilized to evaluate the model for
%Parameter Set 1.
} \end{center}
\end{minipage}
\end{center}

\vfill\eject	 
\begin{center}
\vspace*{16.5cm}
%\special{psfile=anal2.eps hoffset=-20 voffset=0 hscale=67 vscale=60 }
\begin{minipage}[t]{11cm}
\begin{center} {\small {\bf Fig.~2.}
%Simultaneous results for particle spectra and HBT radius parameters.
%The analytic approximations were utilized to evaluate the model for
%Parameter Set 2.
} \end{center}
\end{minipage}
\end{center}

\vfill \eject	 
\begin{center}
\vspace*{16.5cm}
%\special{psfile=num1.eps hoffset=-20 voffset=0 hscale=67 vscale=60}
\begin{minipage}[t]{11cm}
\begin{center} {\small {\bf Fig.~3.}
%Simultaneous results for particle spectra and HBT radius parameters.
%The numeric approximations were utilized to evaluate the model for
%Parameter Set 1.
} \end{center}
\end{minipage}
\end{center}

\vfill \eject	 
\begin{center}
\vspace*{16.5cm}
%\special{psfile=num2.eps hoffset=-20 voffset=0 hscale=67 vscale=60}
\begin{minipage}[t]{11cm}
\begin{center} {\small {\bf Fig.~4.}
%Simultaneous results for particle spectra and HBT radius parameters.
%The numeric approximations were utilized to evaluate the model for
%Parameter Set 2.
} \end{center}
\end{minipage}
\end{center}

\vfill \eject
\begin{center}
\vspace*{16.5cm}
%\special{psfile=err1.eps hoffset=-20 voffset=0 hscale=67 vscale=60}
\begin{minipage}[t]{11cm}
\begin{center} {\small {\bf Fig.~5.}
%Relative deviations between the analytic and the numeric approximations
%to the HBT parameters and particle spectra
%        for Parameter Set 1.
} \end{center}
\end{minipage}
\end{center}

\vfill \eject
\begin{center}
\vspace*{16.5cm}
%\special{psfile=err2.eps hoffset=-20 voffset=0 hscale=67 vscale=60}
\begin{minipage}[t]{11cm}
\begin{center} {\small {\bf Fig.~6.}
%Relative deviations between the analytic and the numeric approximations 
%to the HBT parameters and particle spectra
%        for Parameter Set 2.
} \end{center}
\end{minipage}
\end{center}

\vfill \eject
\begin{center}
\vspace*{16.5cm}
%\special{psfile=small1.eps hoffset=-20 voffset=0 hscale=67 vscale=60}
\begin{minipage}[t]{11cm}
\begin{center} {\small {\bf Fig.~7.}
%Small expansion parameters of the hydrodynamic core model
%        for Parameter Set 1.
} \end{center}
\end{minipage}
\end{center}

\vfill \eject
\begin{center}
\vspace*{16.5cm}
%\special{psfile=small2.eps hoffset=-20 voffset=0 hscale=67 vscale=60}
\begin{minipage}[t]{11cm}
\begin{center} {\small {\bf Fig.~8.}
%Small expansion parameters of the hydrodynamic core model
%        for Parameter Set 2.
} \end{center}
\end{minipage}
\end{center}

\vfill \eject
\begin{center}
\vspace*{16.5cm}
%\special{psfile=errpar1.eps hoffset=-20 voffset=0 hscale=67 vscale=60}
\begin{minipage}[t]{11cm}
\begin{center} {\small {\bf Fig.~9.}
%Parametrization of the relative deviations between the analytic and the 
%numeric approximations to the HBT parameters and particle spectra
%        for Parameter Set 1.
} \end{center}
\end{minipage}
\end{center}

\vfill \eject
\begin{center}
\vspace*{16.5cm}
%\special{psfile=errpar2.eps hoffset=-20 voffset=0 hscale=67 vscale=60}
\begin{minipage}[t]{11cm}
\begin{center} {\small {\bf Fig.~10.}
%Parametrization of the relative deviations between the analytic and the 
%numeric approximations to the HBT parameters and particle spectra
%        for Parameter Set 2.
} \end{center}
\end{minipage}
\end{center}

\vfill \eject
\begin{center}
\vspace*{16.5cm}
%\special{psfile=errpart1.eps hoffset=-20 voffset=0 hscale=67 vscale=60}
\begin{minipage}[t]{11cm}
\begin{center} {\small {\bf Fig.~11.}
%Parametrization of the relative deviations between the analytic and the 
%numeric approximations to the HBT parameters and particle spectra
%        for Parameter Set 1.
} \end{center}
\end{minipage}
\end{center}

\vfill \eject
\begin{center}
\vspace*{16.5cm}
%\special{psfile=errpart2.eps hoffset=-20 voffset=0 hscale=67 vscale=60}
\begin{minipage}[t]{11cm}
\begin{center} {\small {\bf Fig.~12.}
%Parametrization of the relative deviations between the analytic and the 
%numeric approximations to the HBT parameters and particle spectra
%        for Parameter Set 2.
} \end{center}
\end{minipage}
\end{center}

\end{document}